\documentclass[10pt,twocolumn,english]{revtex4}
\usepackage[T1]{fontenc}
\usepackage[latin9]{inputenc}
\usepackage[a4paper]{geometry}
\usepackage{amsmath}
\usepackage{amssymb}
\usepackage{graphicx}

\makeatletter

\newcommand{\be}{\begin{equation}}
\newcommand{\ee}{\end{equation}}
\newcommand{\bi}{\begin{itemize}}
\newcommand{\ei}{\end{itemize}}
\newcommand{\bea}{\begin{eqnarray}}
\newcommand{\eea}{\end{eqnarray}}

\newcommand{\cR}{\mathcal R}
\newcommand{\p}{\partial}
\newcommand{\tg}{\tilde{g}}

\@ifundefined{textcolor}{}
{%
 \definecolor{BLACK}{gray}{0}
 \definecolor{WHITE}{gray}{1}
 \definecolor{RED}{rgb}{1,0,0}
 \definecolor{GREEN}{rgb}{0,1,0}
 \definecolor{BLUE}{rgb}{0,0,1}
 \definecolor{CYAN}{cmyk}{1,0,0,0}
 \definecolor{MAGENTA}{cmyk}{0,1,0,0}
 \definecolor{YELLOW}{cmyk}{0,0,1,0}
}

\makeatother

\usepackage{babel}

\begin{document}

\title{Asymptotic freedom in Ho\v{r}ava-Lifshitz gravity}

\author{Giulio D'Odorico}
\author{Frank Saueressig}
\author{Marrit Schutten}
%\email{x@x} 

%\email{x@x}

\address{
Radboud University Nijmegen, Institute for Mathematics, Astrophysics and Particle Physics,
 Heyendaalseweg 135, 6525 AJ Nijmegen, The Netherlands}

\begin{abstract}
We use the Wetterich equation for foliated spacetimes to study the RG flow of projectable Ho\v{r}ava-Lifshitz gravity coupled to $n$ Lifshitz scalars. Using novel results for anisotropic heat kernels, the matter-induced beta functions for the gravitational couplings are computed explicitly. The RG flow exhibits an UV attractive anisotropic Gaussian fixed point
where Newton's constant vanishes and the extra scalar mode decouples. This fixed point ensures that the theory is asymptotically free in the large--$n$ expansion, indicating that projectable Ho\v{r}ava-Lifshitz gravity is perturbatively renormalizable. Notably, the fundamental fixed point action does not obey detailed balance.
\end{abstract}

\maketitle

\paragraph*{Introduction.}
Classical general relativity provides an excellent description for gravitational phenomena
ranging from sub-millimeter up to astrophysical scales \cite{Will:2014kxa}. Despite these
remarkable successes a fundamental theory capturing the gravitational force on all length scales
is still missing: quantizing the Einstein-Hilbert action along the same lines as QCD
shows that the theory is perturbatively non-renormalizable \cite{'tHooft:1974bx}. In
terms of Wilson's modern formulation of renormalization \cite{Wilson:1973jj}, this feature 
is equivalent to the observation that the Gaussian fixed point (GFP), representing the
free theory, does not act as an UV-attractor for the renormalization
group (RG) flow of Newton's constant $G$, a fact that is easily deduced from the negative mass dimension of $G$. This
raises the question if there are other fixed points of the RG flow which may serve
as a (non-perturbative) UV completion of gravity.

One proposal along these lines is the asymptotic safety program, initially
advocated by Weinberg \cite{Weinberg}. This scenario considers
all action functionals build from the metric field and preserving
(background) diffeomorphism invariance. By now, there is substantial
evidence that the theory space spanned by these action functionals
possesses a non-Gaussian fixed point (NGFP) with a finite number of relevant
deformations \cite{Reuter:2012id}. Thus it is conceivable that gravity 
constitutes a consistent and predictive quantum theory within the framework of non-perturbatively
renormalizable quantum field theories.

Ho\v{r}ava-Lifshitz (HL) gravity \cite{Horava:2008ih,Horava:2009uw} (see \cite{Mukohyama:2010xz,Horava:2011gd,Visser:2011mf,Sotiriou:2009bx}
 for reviews) constitutes an alternative to this scenario. The
construction performs an ADM-decomposition of the $D=d+1$ dimensional 
metric, encoding the gravitational degrees of freedom in the lapse function $N$,
a shift vector $N_i$ and a metric on the $d$-dimensional spatial slices $\sigma_{ij}$,
thereby introducing a foliation structure on spacetime. As its key ingredient
HL gravity relaxes the symmetry requirements underlying asymptotic safety and considers
action functionals invariant with respect to foliation preserving diffeomorphisms only.
%The theory space underlying asymptotic safety therefore constitutes a subspace
%of the one underlying HL gravity.

The reduced symmetry group of HL gravity allows to replace the relativistic flat-space scaling by a scaling relation
that is anisotropic between space and time
\begin{equation}
t \to b \, t, \,\,\,\,\,\, x \to b^{1/z} \, x \, .
\end{equation}
This modifies the standard dimensional analysis and for $z > 1$ interactions including higher powers of spatial derivative terms 
become power-counting relevant or marginal, thereby improving the UV behavior of the theory.
In particular at criticality, $z=d$, Newton's constant is dimensionless so that the theory is power-counting renormalizable
\cite{Horava:2009uw}. Thus it is conceivable that HL gravity is perturbatively renormalizable (asymptotically free). In the language of the Wilsonian renormalization group this conjecture
entails the existence of an anisotropic Gaussian Fixed Point (aGFP) which acts as a UV attractor for Newton's constant at high energies. In \cite{Horava:2009uw}
it was speculated that the aGFP may be located at the conformal point at which the extra scalar mode intrinsic of HL gravity decouples and, in addition, preserve detailed balance.
Given that the aGFP may constitute a valid UV completion of gravity, which is supported by recent Monte Carlo simulations \cite{Ambjorn:2010hu,Ambjorn:2013joa,Anderson:2011bj,Ambjorn:2014gsa}, and the considerable attention devoted to HL gravity,  it is rather surprising that its properties remained largely mysterious and only partial results are available \cite{Anselmi:2007ri,Orlando:2009en,Iengo:2009ix,Giribet:2010th,Contillo:2013fua,Benedetti:2013pya}. 

This situation clearly calls for the use of the RG to clarify the structure of HL gravity at high energies.
A convenient way to perform such computations 
is the functional RG equation for the gravitational effective average action $\Gamma_k$ \cite{Reuter:1996cp}.
The metric construction can readily be adapted to the case of projectable Ho\v{r}ava-Lifshitz (pHL) gravity \cite{Rechenberger:2012dt},
yielding the Wetterich equation governing the scale-dependence of $\Gamma_k$ 
\be\label{FRGE}
\partial_t \Gamma_k = \tfrac{1}{2} \, {\rm Tr} \left[ \left(\Gamma_k^{(2)} + \cR_k \right)^{-1} \partial_t \cR_k \right] \, . 
\ee
Here $t \equiv \ln(k/k_0)$ is the ``RG time'', $\Gamma_k^{(2)}$ is the Hessian of $\Gamma_k$, and $\cR_k$ is a regulator term designed to suppress low energy modes.
In the sequel, we will work with the Litim regulator \cite{Litim:2001up}, setting $\cR_k(y) = (k^2 - y) \theta(k^2-y)$.

In this letter, we use the flow equation \eqref{FRGE} to study the ultraviolet behavior of pHL gravity coupled to $n$ anisotropic
Lifshitz scalar fields in the large--$n$ limit. In this limit the beta functions are dominated by the contributions originating
from integrating out the scalar fields. Since gravity couples universally to itself as well as to matter, the contribution of the graviton propagating in loops 
becomes negligible with respect to the contribution of the $n$ matter fields 
and the beta functions become exact in the large $n$-limit. 
%so that the the contribution from gravity loops may be neglected. 
As we will show, this ansatz allows us to identify the aGFP underlying the perturbative renormalizability of HL gravity in $D=3+1$ dimensions.

\paragraph*{Setup.}
Concretely, the effective average action of the system is given by 
projectable Ho\v{r}ava--Lifshitz gravity coupled to $n$ Lifshitz scalars
%{\bf Ansatz from Frank:}
%In order to study the high-energy behavior of the theory we consider 
%
\be\label{eaa}
\Gamma_k[N,N_i,\sigma, \phi] = \Gamma_k^{\rm HL}[N,N_i,\sigma] + S^{\rm LS}[N,N_i,\sigma, \phi] \, . 
\ee
Here
\be
S^{\rm LS} \equiv \tfrac{1}{2}\int dtd^{d}xN\sqrt{\sigma}\phi\left[\Delta_{t}+(\Delta_{x})^{z}\right]\phi
\ee
is the action of the Lifshitz scalars with anisotropic scaling $z$ in a curved background, 
%
%\be
$\Delta_{t} \equiv - \tfrac{1}{N\sqrt{\sigma}} \, \p_t \, N^{-1} \sqrt{\sigma} \, \p_t$,
%\ee
%
and $\Delta_x \equiv - \sigma^{ij} D_i D_j$ is the positive definite covariant Laplacian on the spatial slice
with $D_i$ containing the $d$-dimensional Christoffel symbols constructed from $\sigma_{ij}(t,x)$.
The gravitational part of $\Gamma_k$ is given by the classical action for pHL
 gravity \cite{Horava:2009uw}
\be
\Gamma_k^{\rm HL} = \tfrac{1}{16\pi G_k}  \int dtd^{d}xN\sqrt{\sigma} \left[ K_{ij}K^{ij}-\lambda_k K^{2} + V_k \right]
\ee
where all couplings $g_i(k)$ have been promoted to depend on the RG scale $k$.
The potential $V_k[\sigma]$ contains all power-counting relevant and marginal operators
constructed from the intrinsic curvature tensors on the spatial slices.
Following \cite{Sotiriou:2009bx}, $V_k$ for $d=2$ has the form 
\be\label{pot2d}
V_k^{(d=2)} = g_0 + g_1 \, R + g_2 \, R^2 \, ,
\ee
while the case $d=3$ includes all interactions with up to six spatial derivatives
\be\label{pot3d}
\begin{split}
& \, V_k^{(d=3)} =  g_{0}+g_{1}R+g_{2}R^{2}+g_{3}R_{ij}R^{ij}- g_{4}R \Delta_x R \\ & \;  - g_{5}R_{ij} \Delta_x R^{ij} 
 + g_{6}R^{3}+g_{7}RR_{ij}R^{ij}+g_{8}R_{j}^{i}R_{k}^{j}R_{i}^{k} \, . 
\end{split}
\ee
 Thus the setup contains two wave-function renormalizations $(G_k, \lambda_k)$ and  
three ($d=2$) and eight ($d=3$) running parameters in the potential, respectively.
Our conventions for parameterizing $V_k$ in terms of coupling constants follow 
the ones typically adopted in higher-derivative gravity
 and are tailored to exhibit the properties of potential fixed points.
Note that we have not implemented detailed balance to simplify $V_k$.

\paragraph*{Beta functions.}
We now compute the beta functions, $\beta_{g_i} \equiv \p_t g_i$, capturing
the running of the gravitational coupling constants induced by integrating
out the scalar fields. For this purpose
it is convenient to express the flow in terms of dimensionless couplings 
$\tg_i \equiv g_i k^{-[g_i]}$, where
$[g_i]$ is the canonical mass dimension of $g_i$. In particular
the dimensionless Newton's constant is given by
\be\label{dimnewton}
g_k \equiv G_k \, k^{2\eta} \, , \qquad \eta \equiv \tfrac{d}{2z}-\tfrac{1}{2} \, ,  
\ee
so that $G_k$ becomes dimensionless at criticality $d=z$. Actually it is
the shift of the canonical mass dimension induced by the anisotropy $z$
which underlies the power-counting renormalizability of HL gravity.

The beta functions are obtained by substituting \eqref{eaa} into \eqref{FRGE}
and reading off the coefficients multiplying the extrinsic and intrinsic 
curvatures. In the case where  the RG flow
is driven by the scalar fields, the operator trace is given
by ${\rm Tr} \, W(\square)$ with $W(y) = (y + R_k(y))^{-1} \p_t R_k(y)$
and $\square \equiv  \Delta_t + \left( \Delta_x \right)^z$ being
the anisotropic Laplace operator.

 The evaluation of the trace
can be linked to the short-time expansion of
the heat kernel ${\rm Tr}\, e^{-s \square}$ 
by re-expressing $W(z)$ through its Laplace transform. 
The expansion of the heat kernel can be found in a systematic way by the repeated use of the
 Baker--Hausdorff Lemma combined with the off-diagonal heat-kernel techniques \cite{Anselmi:2007eq,Benedetti:2010nr}.
 The result is given by
\begin{equation}
\begin{split}
{\rm Tr} & \, e^{-s \square} \simeq   \left(4\pi \right)^{-(d+1)/2}  s^{-(1+d/z)/2} \, \int dt d^dx N \sqrt{\sigma} \, \times \\
& \, \bigg[ \frac{s}{6} \left( e_1 \, K^2 + e_2 \, K_{ij} K^{ij} \right) + \sum_{n \ge 0 } \, s^{n/z} \,  b_n \, a_{2n} \, \bigg]   \, .
\end{split}
\label{master}
\end{equation}
Here $a_{2n}$ are the standard heat-kernel coefficients containing the intrinsic curvatures
constructed from $\sigma_{ij}$, possibly subject to geometrical constraints arising from
working on a low-dimensional manifold. Their values have, e.g., been given in \cite{Avramidi:2000pia,Groh:2011dw}: 
$a_0 = 1$, $a_2 = R/6$, etc. The coefficients $e_i$ and $b_n$ depend on $d,z$ and encode the corrections originating from the anisotropic differential operator. The 
ones in the extrinsic curvature sector are
\be\label{ee}
e_1 = \frac{d-z+3}{d+2} \frac{\Gamma(\tfrac{d}{2z})}{z \Gamma(\tfrac{d}{2})} \, , \quad
e_2 = - \frac{d+2z}{d+2} \frac{\Gamma(\tfrac{d}{2z})}{z \Gamma(\tfrac{d}{2})} \, , 
\ee
while the $b_n(d,z)$ for $0 \le n \le \lfloor d/2 \rfloor$ are given by
\be\label{bis}
b_n = \frac{\Gamma\left( \tfrac{d-2n}{2z} +1\right)}{\Gamma\left( \tfrac{d-2n}{2} + 1 \right)} \, .
\ee
The coefficients $b_n(d,z)$ for $n >  \lfloor d/2 \rfloor$ can be computed 
on a case by case basis. The ones relevant for the present computation are 
\be
b_2(2,2) = 0 \, , \;\;  b_2(3,3) = \tfrac{1}{\sqrt{\pi}} \Gamma(\tfrac{5}{6}) \, , \; \; b_3(3,3) = - \tfrac{1}{2} \, .
\ee
For $z=1$ these results reproduce the standard heat kernel written in terms of ADM-variables 
while for $d=z=2$ our formulas coincide with the special case considered in \cite{Baggio:2011ha}.  
The systematic computation of the early-time expansion of the heat kernel of an anisotropic Laplace
operator constitutes the main technical breakthrough of our work.

Based on the expansion \eqref{master}, it is straightforward to obtain
the desired beta functions by applying the Mellin transform techniques reviewed, e.g., in \cite{Codello:2008vh}. 
Defining 
%
%\be
$\phi_n \equiv \tfrac{1}{\Gamma(n)} \, \int_0^1 \, dx \, x^{n-1}$,
%\ee
%
the scale-dependence of the two (dimensionless) wave-function renormalizations is governed by
\be\label{wfctren}
\begin{split}
\beta_g = & \, 2 \, \eta \, g - \tfrac{2}{3} \left(4\pi\right)^{-(d-1)/2} \, \phi_{\eta} \, e_2 \,  g^2 \, , \\
\beta_\lambda = \, & - \tfrac{2}{3} \left(4\pi\right)^{-(d-1)/2}  \, \phi_{\eta} \, \left( e_1 + \lambda \, e_2 \right) \, g \, ,
\end{split}
\ee
while the beta functions for the cosmological constant $\tg_0$ and for $\tg_1$ read
\be\label{ehren}
\begin{split}
\beta_{\tg_0} = & \, -2 \, \tg_0 + \tfrac{4g}{(4\pi)^{(d-1)/2}} \, \left( b_0 \, \phi_{\eta+1} - \tfrac{1}{6} \, e_2 \, \phi_{\eta} \, \tg_0 \right) , \\
\beta_{\tg_1} = \, & \left( \tfrac{2}{z}-2\right) \tg_1 +  \tfrac{2g}{3 (4\pi)^{(d-1)/2}} \, \left( b_1 \, \phi_{\eta+1-1/z} - e_2 \, \phi_{\eta} \, \tg_1 \right) \, .
\end{split}
\ee
At criticality $d=z=2$ the system is supplemented by
\be
\beta_{\tg_2} = \tfrac{g}{4} \, \tg_2 \, .
\ee
For $d=z=3$ the additional beta functions read
\be
\begin{split}
\beta_{\tg_2} = & \, - \tfrac{2}{3} \tg_2 + \tfrac{g}{5\pi} \left( \tfrac{1}{8\sqrt{\pi}} \tfrac{\Gamma(5/6)}{\Gamma(1/3)} +  \tg_2 \right) \, , \\
\beta_{\tg_3} = & \, - \tfrac{2}{3} \tg_3 + \tfrac{g}{5\pi} \left( \tfrac{1}{4\sqrt{\pi}} \tfrac{\Gamma(5/6)}{\Gamma(1/3)} +  \tg_3 \right) \, , \\
\beta_{\tg_i} = & \tfrac{g}{\pi} \left( \tfrac{1}{5} \, \tg_i - \tfrac{1}{2} \, c_i \right) \, , \; \; \; i = \{4,5,6,7,8\} \, , 
\end{split}
\label{d3marg}
\ee
with the  numerical constants
\be
\renewcommand{\arraystretch}{1.5}
\begin{array}{lll}
c_4 = \tfrac{1}{336} \, , \; &  c_5 = \tfrac{1}{840} \, , & \\
c_6 = - \tfrac{1}{560} \, , \; & c_7 = \tfrac{1}{105} \, , \; & c_8 = - \tfrac{1}{180} \, . 
\end{array}
\ee

We now summarize the most important properties of these flow equations.

\begin{figure}
\begin{centering}
\includegraphics[scale=0.7]{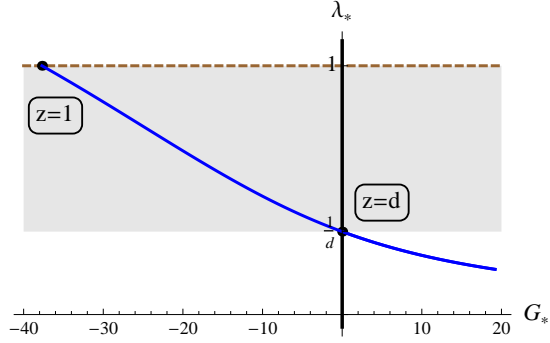}
\par
\end{centering}
\caption{
$z$--dependence of the fixed point arising from (\ref{wfctren}). The dashed horizontal line represents the general relativistic case, while the thick vertical line represents the line of Gaussian fixed points. In the shaded area the extra scalar mode of HL gravity is classically unstable \cite{Visser:2011mf}.}
\label{fps}
\end{figure}
\emph{Isotropy: $z=1$.} in the isotropic case eqs.\ \eqref{wfctren} and \eqref{ehren} exhibit a fixed plane
$\{ \lambda^* = 1, \tg_1^* = - 1\}$ which is invariant under RG transformations.
This plane possesses a NGFP situated at
\be\label{NGMF}
\begin{split}
\tg_0^* =  - \tfrac{12(d-1)}{(d+1)^2} \, , \; 
g^* =  - \tfrac{3(d-1)}{2} (4\pi)^{(d-1)/2} \, \Gamma\left( \tfrac{d+1}{2} \right) \, . 
\end{split}
\ee
For $d=3$ this fixed point coincides with the one found in the metric computations
\cite{Codello:2008vh,Dona:2013qba}. On the fixed plane the two-derivative terms of 
$\Gamma_k^{\rm HL}$ combine into the $d+1$-dimensional Einstein-Hilbert action. 
Thus a flow starting on this subspace will not generate Lorentz violating interactions.
This result provides first hand evidence that gravitational interactions preserving full
diffeomorphism invariance span a subspace in the theory space underlying HL gravity which is closed under RG flows.

 \emph{ Criticality: $z=d$.} For $z > 1$ the NGFP \eqref{NGMF} is shifted
 towards smaller values $\lambda^* < 1$, see Fig.\ \ref{fps}. At criticality
 $\eta = 0$, $e_2 = - d e_1$ so that the fixed point is located at
 \be\label{aGFP}
{\rm aGFP:} \qquad g^* = 0 \, , \quad \lambda^* = \tfrac{1}{d} \, . 
 \ee
In $d=2$ the aGFP is completed by
\be\label{aGFP2}
\tg_1 ^* = 0 \, , \quad \tg_2 ^* = 0 \, , \quad \tg_3 ^* = 0
\ee
while in $d=3$ the couplings determining $V_*$ are fixed to
\be\label{aGFP3}
\tg_1 ^* = \tg_2 ^* = \tg_3 ^* = 0 \, , \qquad \tg_i^*  = \frac{5}{2} c_i \, . 
\ee
Newton's constant vanishes at the aGFP justifying the label ``Gaussian''. Moreover,
the fixed point is precisely situated at the point where the extra scalar mode becomes
non-dynamical. It is this $z=d$ Lifshitz fixed point that underlies the conjectured
perturbative renormalizability of HL gravity. 

\begin{figure}
\begin{centering}
\includegraphics[scale=0.67]{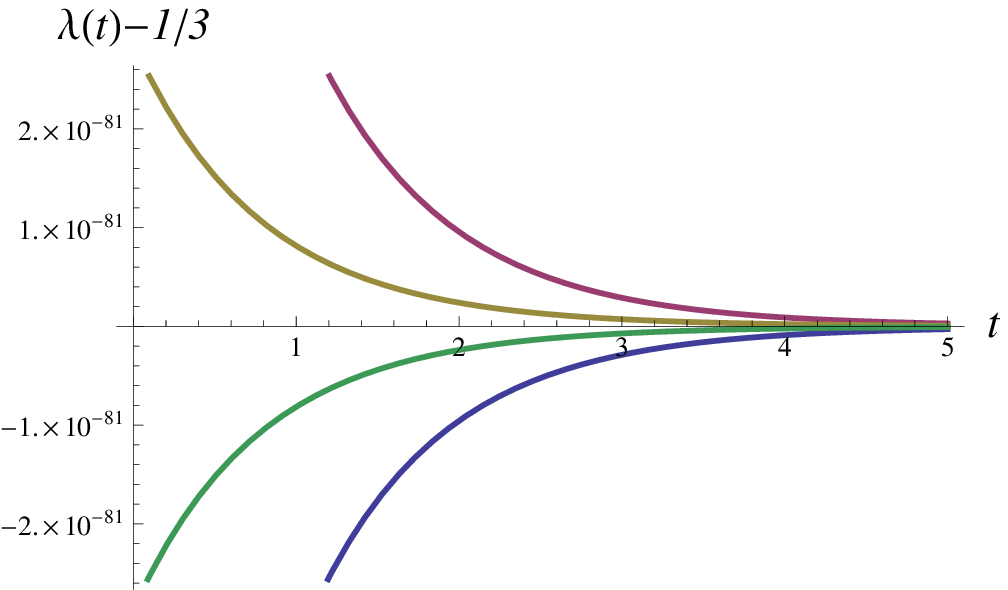}
\includegraphics[scale=0.67]{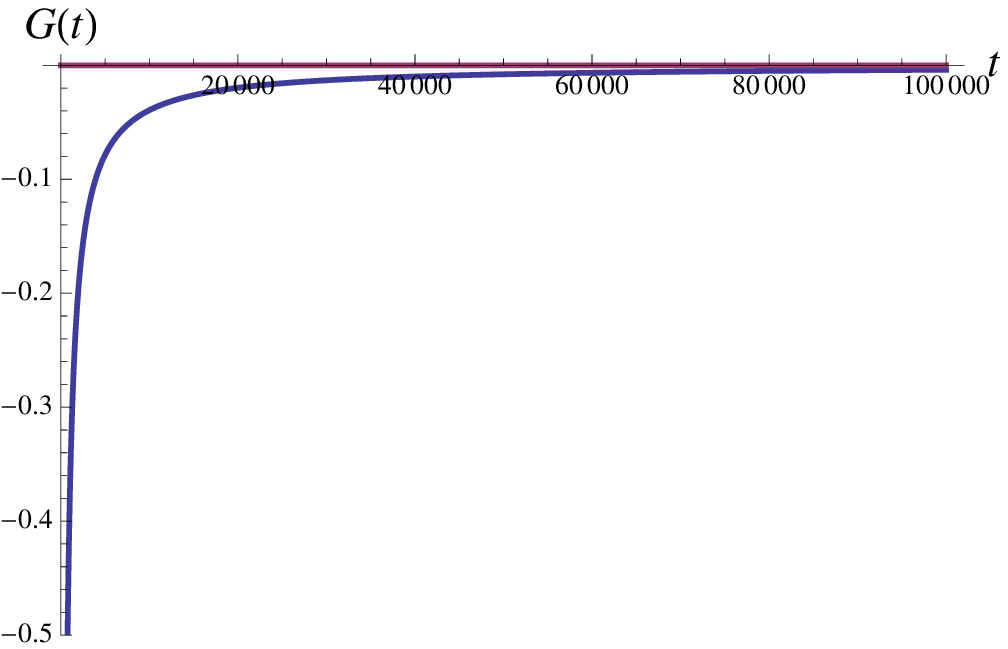}\\
\includegraphics[scale=0.67]{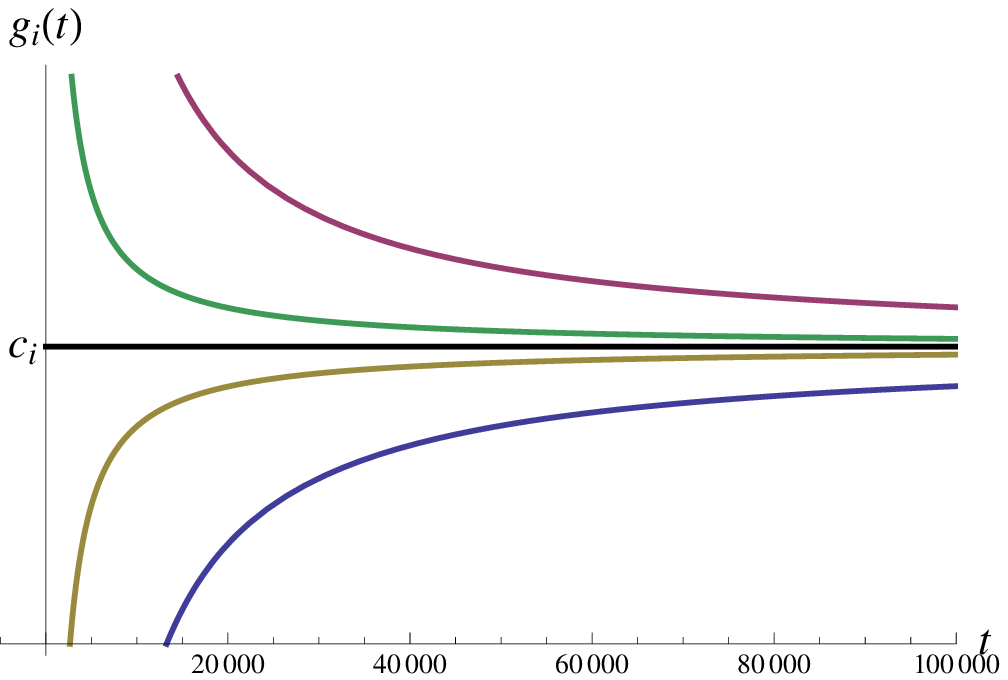}
\par
\end{centering}
\caption{
Scale-dependence of the wave-function renormalizations $G_k$ and $\lambda_k$ and the marginal couplings $\tg_i$ in $d=z=3$.
The aGFP (\ref{aGFP},\ref{aGFP3}) acts as a UV attractor of the RG flow.}
\label{agfp}
\end{figure}
The scale-dependence of the coupling constants at criticality $z=d=3$ is
shown in Fig.\ \ref{agfp}. Clearly, the aGFP acts as a UV attractor of the
RG flow, supporting the hypothesis that the theory is asymptotically free.
Somewhat intriguing, $G_k$ flows to zero for $G_k < 0$ and the regime $G_k > 0$ is
separated from the aGFP by a Landau pole. Based on the analogous computations
within the asymptotic safety program \cite{Codello:2008vh,Dona:2013qba}, where the inclusion of gravity
loops shifts the matter induced NGFP from $g^* < 0$ to $g^* >0$, we
expect that this feature will be cured by taking gravitational fluctuations
into account. We hope to come back to this point in the future. For the time
being we will content ourselves with discussing two eminent consequences of our findings.

%%%%%%%%%%%%%%%%%%%%%
\paragraph*{Flat space propagators at criticality.}
%%%%%%%%%%%%%%%%%%%%%
It is illustrative to expand the $3+1$-dimensional fixed point action $\Gamma_*^{\rm HL}$ 
obtained from the aGFP around (Euclidean) flat space. Since the scalar mode 
is non-dynamical at the aGFP we focus on the transverse traceless (TT) fluctuations
only (also see \cite{Benedetti:2013pya} for a more detailed discussion). Expanding $\sigma_{ij} = \delta_{ij} + \epsilon h_{ij}$ with
$\epsilon=\sqrt{16\pi g_*}$ yields the two-point propagator
\be\label{prop}
{\cal G}_*^{\rm TT} \propto \left( \omega^2 - \tg_5^* \, \vec{p}^{\,6} \right)^{-1} \, ,
\ee
where $\omega$ and $\vec{p}$ denote the energy and spatial momentum of the graviton, respectively.
The higher order vertices vanish at the aGFP. Thus the aGFP describes a non-interacting theory
with an $z=3$ anisotropic dispersion relation. At this stage the following remarks are in order.
The sign of $\tg_5^*$ indicates that the propagator is unstable in Euclidean space. Moreover, performing
the same analysis in $d=2$  shows that the propagator contains only an energy term, owed to the vanishing of $\tg_2^*$.
We expect that the inclusion of gravity loops will resolve these issues. Our results
then indicate the existence of an upper bound on the number of scalar fields that can
consistently be coupled to pHL gravity without destabilizing the gravitational sector.

%%%%%%%%%%%%%
\paragraph*{Detailed balance.}
%%%%%%%%%%%%% 
A conjecture already put forward in the seminal work \cite{Horava:2009uw} is that
the potentials (\ref{pot2d},\ref{pot3d}) satisfy detailed balance,
stating that $V$ can be derived from a variational principle,
%
%\be
$V\propto \frac{\delta W\left[ \sigma \right]}{\delta \sigma_{ij}} {H}_{ijkl} \frac{\delta W\left[ \sigma \right]}{\delta \sigma_{kl}} $ %\, , 
%\ee
%
with ${H}$ being the de Witt supermetric. For $z=d=3$ this conjecture implies
that the superpotential $W$ generating the six-derivative terms is the Chern-Simons action, so that the potential is given by the square of the Cotton tensor, 
 $ V_{\rm db} \propto C_{ij}C^{ij}$, with 
 $C^{ij} = \epsilon^{ikl} D_k \left(R^j_l - \tfrac{1}{4} R \delta^j_l\right)$.
Rewriting $ V_{\rm db}$ in terms of the basis \eqref{pot3d} yields
 \be
 \begin{split}
  V_{\rm db} \propto & \,   R_{ij} \Delta_x R^{ij} - \tfrac{3}{8} R \Delta_x R  \\ & \, + 3 R^i_j R^j_k R^k_i - \tfrac{5}{2} R R^{ij}R_{ij} + \tfrac{1}{2} R^3 \, . 
 \end{split}
 \ee
Constructing $V_*$ from the fixed point couplings  \eqref{aGFP3} shows that
the scalar induced aGFP \emph{does not obey detailed balance}. This feature is actually independent of the
correction coefficients $b_n$ and solely relies on the sixth order heat-kernel coefficients $a_6$.

%%%%%%%%%%%%%
\paragraph*{Conclusions.}
%%%%%%%%%%%%%
In this letter, we used the functional renormalization group equation for projectable Ho\v{r}ava-Lifshitz (pHL) gravity \cite{Rechenberger:2012dt},
to study the matter-induced RG flow of the gravitational coupling constants in the large--$n$ limit. 
As a key result we identified the $z=d$ Lifshitz critical point (aGFP) underlying the perturbative renormalizability of the theory
and showed that it acts as a UV attractor for the RG flow. The aGFP is situated at conformal point where the extra degree of 
freedom intrinsic to HL gravity is non-dynamical and the underlying fixed point action does not preserve detailed balance.
While the computation has been carried out in the framework of pHL gravity, we expect
that the aGFP also acts as UV fixed point for the non-projectable theory, the difference being that, owed to the enlarged theory space, the latter
case will give rise to a larger number of relevant deformations.
Clearly, it is interesting to complement the present computation by the inclusion of gravity loops and we hope to come back to this point in the near future.\\

\paragraph*{Acknowledgements.}
We thank A.\ Contillo, S.\ Gryb and S.\ Rechenberger for helpful discussions.
The research of F.~S.\ and G.~D.\ is supported by the Deutsche Forschungsgemeinschaft (DFG)
within the Emmy-Noether program (Grant SA/1975 1-1).

\end{document}